\documentstyle[12pt,epsfig]{article}

\def\baselinestretch{1.2}
\def\href#1#2{#2}  

\newcommand{\norm}[1]{\raise.3ex\hbox{:} #1 \raise.3ex\hbox{:}\,}

\newfont{\Bbb}{msbm10 scaled 1200}     
\newcommand{\mathbb}[1]{\mbox{\Bbb #1}}

\def\cM{{\cal M}}
\def\cN{{\cal N}}
\def\cF{{\cal F}}

\newcommand{\beq}{\begin{equation}}
\newcommand{\eeq}{\end{equation}}
\newcommand{\beqar}{\begin{eqnarray}}
\newcommand{\eeqar}{\end{eqnarray}}
\def\appendix{{\newpage\section*{Appendix}}\let\appendix\section%
        {\setcounter{section}{0}
        \gdef\thesection{\Alph{section}}}\section}

\newcommand{\be}{\begin{equation}}
\newcommand{\ee}{\end{equation}}
\newcommand{\eel}[1]{\label{#1}\end{equation}}
\newcommand{\bea}{\begin{eqnarray}}
\newcommand{\eea}{\end{eqnarray}}
\newcommand{\eeal}[1]{\label{#1}\end{eqnarray}}
\newcommand{\baq}{\begin{equation}\begin{array}{rcl}}
\newcommand{\eaq}{\end{aryray}\end{equation}}
\newcommand{\eaql}[1]{\end{array}\label{#1}\end{equation}}
\newcommand{\beac}{\begin{equation}\begin{array}{rcl}}
\newcommand{\eeacn}[1]{\end{array}\label{#1}\end{equation}}
\newcommand{\ba}{\begin{array}}
\newcommand{\ea}{\end{array}}

\newcommand{\journal}[4]{{\rm #1~}{\bf #2}\,(#3)\,#4}

\newcommand{\ijmp}{\journal {Int. J. Mod. Phys.}}

\newcommand{\pr}{\journal {Phys. Rev.}}

\newcommand{\prl}{\journal {Phys. Rev. Lett.}}

\newcommand{\rmp}{\journal {Rev. Mod. Phys.}}

\newcommand{\cmp}{\journal {Comm. Math. Phys.}}
\newcommand{\cqg}{\journal {Class. Quantum Grav.}}

\newcommand{\np}{\journal {Nucl. Phys.}}
\newcommand{\pl}{\journal {Phys. Lett.}}
\newcommand{\mpl}{\journal {Mod. Phys. Lett.}}

\newcommand{\ptp}{\journal {Progr. Theor. Phys.}}
\newcommand{\nc}{\journal {Nuovo Cim.}}

\newcommand{\grg}{\journal {Gen. Rel. Grav.}}

\catcode`\@=11

\@addtoreset{equation}{section}
\catcode`\@=12

\def\noj#1,#2,{{\bf #1} (19#2)\ }
\def\jou#1,#2,#3,{{\sl #1\/ }{\bf #2} (19#3)\ }
\def\ann#1,#2,{{\sl Ann.\ Physics\/ }{\bf #1} (19#2)\ }
\def\cmp#1,#2,{{\sl Comm.\ Math.\ Phys.\/ }{\bf #1} (19#2)\ }
\def\ma#1,#2,{{\sl Math.\ Ann.\/ }{\bf #1} (19#2)\ }
\def\jd#1,#2,{{\sl J.\ Diff.\ Geom.\/ }{\bf #1} (19#2)\ }
\def\invm#1,#2,{{\sl Invent.\ Math.\/ }{\bf #1} (19#2)\ }
\def\cq#1,#2,{{\sl Class.\ Quantum Grav.\/ }{\bf #1} (19#2)\ }
\def\cqg#1,#2,{{\sl Class.\ Quantum Grav.\/ }{\bf #1} (19#2)\ }
\def\ijmp#1,#2,{{\sl Int.\ J.\ Mod.\ Phys.\/ }{\bf A#1} (19#2)\ }
\def\jmphy#1,#2,{{\sl J.\ Geom.\ Phys.\/ }{\bf #1} (19#2)\ }
\def\jams#1,#2,{{\sl J.\ Amer.\ Math.\ Soc.\/ }{\bf #1} (19#2)\ }
\def\grg#1,#2,{{\sl Gen.\ Rel.\ Grav.\/ }{\bf #1} (19#2)\ }
\def\mpl#1,#2,{{\sl Mod.\ Phys.\ Lett.\/ }{\bf A#1} (19#2)\ }
\def\nc#1,#2,{{\sl Nuovo Cim.\/ }{\bf #1} (19#2)\ }
\def\np#1,#2,{{\sl Nucl.\ Phys.\/ }{\bf B#1} (19#2)\ }
\def\pl#1,#2,{{\sl Phys.\ Lett.\/ }{\bf #1B} (19#2)\ }
\def\pla#1,#2,{{\sl Phys.\ Lett.\/ }{\bf #1A} (19#2)\ }
\def\pr#1,#2,{{\sl Phys.\ Rev.\/ }{\bf #1} (19#2)\ }
\def\prd#1,#2,{{\sl Phys.\ Rev.\/ }{\bf D#1} (19#2)\ }
\def\prl#1,#2,{{\sl Phys.\ Rev.\ Lett.\/ }{\bf #1} (19#2)\ }
\def\prp#1,#2,{{\sl Phys.\ Rept.\/ }{\bf #1C} (19#2)\ }
\def\ptp#1,#2,{{\sl Prog.\ Theor.\ Phys.\/ }{\bf #1} (19#2)\ }
\def\ptpsup#1,#2,{{\sl Prog.\ Theor.\ Phys.\/ Suppl.\/ }{\bf #1}
(19#2)\ }
\def\rmp#1,#2,{{\sl Rev.\ Mod.\ Phys.\/ }{\bf #1} (19#2)\ }
\def\yadfiz#1,#2,#3[#4,#5]{{\sl Yad.\ Fiz.\/ }{\bf #1} (19#2) #3%
\ [{\sl Sov.\ J.\ Nucl.\ Phys.\/ }{\bf #4} (19#2) #5]}
\def\zh#1,#2,#3[#4,#5]{{\sl Zh..\ Exp.\ Theor.\ Fiz.\/ }{\bf #1}
(19#2) #3%
\ [{\sl Sov.\ Phys.\ JETP\/ }{\bf #4} (19#2) #5]}

\begin{document}

\begin{titlepage}

\begin{flushright}
CERN-TH/99-229\\
hep-th/9907206
\end{flushright}
\vfil\vfil

\begin{center}

{\Large {\bf AdS/CFT and BPS Strings in Four Dimensions\\
}}

\vfil

Mohsen Alishahiha and Yaron Oz

\vfil

Theory Division, CERN\\
CH-1211, Geneva 23, Switzerland

\end{center}

\vspace{5mm}

\begin{abstract}
\noindent 
We consider $\cN=2$ superconformal
theories defined on a $3+1$ dimensional hyperplane 
intersection of two sets 
of M5 branes.
These theories have (tensionless)
BPS string solitons. We use a dual supergravity
formulation to deduce some of their properties via the AdS/CFT correspondence.

\end{abstract}

\vfil\vfil\vfil
\begin{flushleft}
July 1999
\end{flushleft}
\end{titlepage}

\newpage
\renewcommand{\baselinestretch}{1.05}  

\section{Introduction}

The $\cN=2$ supersymmetry algebra in four dimensions
contains a central extension term that corresponds to string charges
in the adjoint representation of the $SU(2)_R$ part of the R-symmetry group.
This implies that $\cN=2$ supersymmetric gauge theories in four dimensions
can have BPS string configurations at certain regions  in their moduli space
of vacua. 
In particular, at certain points in the moduli space of vacua these strings
can become tensionless.
A brane configuration that exhibits this phenomena consists of two  sets 
of M5 branes intersecting on $3+1$ dimensional hyperplane.
The theory on the intersection is $\cN=2$ supersymmetric.
One can stretch M2 branes between the two sets of M5 branes in a configuration
that preserves
half of the supersymmetry.
This can be viewed as a BPS string of the four dimensional theory.
The purpose of this note is to study such brane configurations using the AdS/CFT 
correspondence \cite{mal,us}.
Such a non-localized brane system has been discussed in \cite{HK},
and it does not correpsond to an $\cN=2$ SCFT. The fully localized
brane system was given a DLCQ description as
a supersymmetric quantum mechanics in \cite{KOY}. This system does lead to
an $\cN=2$ SCFT on the $3+1$ dimensional intersection and will be discussed in the paper.
Most of our detailed analysis will be done for a semi-localized brane system which corresponds 
to an $\cN=2$ SCFT too.

The paper is organized as follows. In the next section we will present the 
dual supergravity description of the four dimensional theory on the 
intersection and discuss the field theory on the intersection.
In section 3 we will use the dual supergravity description to deduce
some properties of these strings. We will argue that from the 
four dimensional field theory viewpoint they are simply BPS string configurations
on the Higgs branch.

\section{The Supergravity Description}

We denote the eleven dimensional space-time coordinates by $(x_{||},\vec{x},\vec{y},\vec{z})$,
where $x_{||}$ parametrize the $(0,1,2,3)$ coordinates,
$\vec{x}=(x_1,x_2)$ the $(4,5)$ coordinates, 
$\vec{y} = (y_1,y_2)$ the $(6,7)$ coordinates and 
$\vec{z}=(z_1,z_2,z_3)$ the $(8,9,10)$ coordinates.

Consider two sets of fivebranes in M theory: $N_1$ coinciding M5 branes and $N_2$ 
coinciding M5' branes.
Their worldvolume coordinates are  $M5: (x_{||},\vec{y})$
and $M5': (x_{||},\vec{x})$. 
Such a configuraion preserves eight supercharges.
The eleven-dimensional supergravity background takes the form \cite{PT,T,GKT}
\beq
ds^2_{11} = (H_1H_2)^{2/3}[(H_1H_2)^{-1}dx^2_{||}
+H^{-1}_2 d\vec{x}^2+H^{-1}_1 d\vec{y}^2+ d\vec{z}^2] \ ,
\label{m5m5}
\eeq
with the 4-form field strength $\cF$
\beq
\cF = 3(*dH_1\wedge dy^1 \wedge dy^2 + *dH_2\wedge dx^1 \wedge dx^2) \ ,
\eeq
where $*$ defines the dual form in the three dimensional space $(z_1,z_2,z_3)$.
When the M5 and M5' branes are only localized along the overall transverse directions $\vec{z}$
the  harmonic functions are $H_1 = 1 + l_p N_1/2|\vec{z}|,H_2 = 1 + l_p N_2/2|\vec{z}|$,
where $l_p$ is the eleven dimensional Planck length.
The near horizon geometry in this case does not have the AdS isometry group and thus cannot
describe a dual SCFT. Moreover, there does not seem to be in this case
a theory on the intersection which decouples from the bulk physics.

Consider the semi-localized case when the M5 branes are completely localized while the
M5' branes are only localized along the overall transverse directions \footnote{
The supergravity solution with both the M5 branes and the M5' branes being completely localized
has not been constructed yet. We will discuss this system later.}.  
When the branes are at the origin of $(\vec{x},\vec{z})$ 
space the harmonic functions in the near core limit of the M5' branes take the form
\cite{Youm,L} \footnote{The numerical factors in (\ref{m5m5harm})
are determined by the requirment that the integral of ${\cal F}$ yields
the appropriate charges.
This solution can also be obtained from the localized D2-D6 brane solution of
\cite{ity} by a chain of dualities. 
} 
\beq
H_1=1+\frac{4\pi l_p^4 N_1 N_2}{(|\vec{x}|^2+
2l_p N_2|\vec{z}|)^2}, \ \ \  
H_2=\frac{l_p N_2}{2|\vec{z}|} \ .
\label{m5m5harm}
\eeq

It is useful to make a change of coordinates
$l_p z= (r^2 \sin\alpha^2)/2N_2, x=r \cos\alpha, 0 \leq \alpha \leq \pi/2$.
In the
near-horizon limit we want to keep the energy $U=\frac{r}{l_p^2}$ fixed.
This implies, in particular, that membranes stretched between the M5 snd M5' branes
in the $\vec{z}$ direction have finite tension $|\vec{z}|/l_p^3$.
Since we are smearing over the $\vec{y}$ directions we should keep $\vec{y}/l_p$ 
fixed. It is useful to make 
a change of variables  $\vec{y}/l_p \rightarrow \vec{y}$.
The near horizon metric is of the form of a $warped\,\,\, product$ of $AdS_5$ 
and a six dimensional manifold $\cM_6$
\beq
ds^2_{11} = l_p^2 (4 \pi N_1)^{-1/3}(\sin^{2/3}\alpha) \left (
\frac{U^2}{N_2}
dx^2_{||}+ \frac{4\pi N_1}{U^2} dU^2+ d\cM_6^2 \right ) \ , 
\label{metric11}
\eeq
with
\beq
d\cM_6^2 = 
4\pi N_1 (d\alpha^2+\cos^2\alpha d\theta^2 
+\frac{\sin^2\alpha}{4}d\Omega_2^2 )
+\frac{N_2}{\sin^2 \alpha}(dy^2+y^2d\psi^2) \ .
\label{M}
\eeq
The metric (\ref{metric11}) has the $AdS_5$ isometry group \cite{van}. Therefore, in the
spirit
of \cite{mal},  M theory on
the background  (\ref{metric11}), (\ref{M})
should be dual to a four dimensional $\cN=2$ SCFT.
Similar types of metrics appear in the near-horizon limit of the localized
D2-D6 \cite{ity,pelc} and D4-D8 brane systems \cite{bo}.

Note that the curvature of the metric diverges for small $\alpha$ as
\beq
{\cal R}\sim \frac{1}{l_p^2N_1^{2/3}
\sin^{8/3}\alpha} \ .
\eeq
Away from $\alpha=0$, eleven-dimensional supergravity can be trusted for large $N_1$.  
The singularity at $\alpha=0$ is interpreted as a signal that some degrees of freedom
have been effectively integrated and are needed in order to resolve the singularity.
These presumably correspond to membranes that end on the M5' branes.


The above M5-M5' brane system can be understood as up lifting to eleven dimensions of an elliptic brane
system of Type IIA. It consists of
$N_2$ NS5-branes with worldvolume coordinates $(0,1,2,3,4,5)$ 
periodically arranged in the 6-direction and 
$N_1$ D4-branes with worldvolume coordinates $(0,1,2,3,6)$ 
stretched between them as in figure 1. 
When we lift this brane configuration to eleven dimensions, we delocalize in the eleven 
coordinate (which in our notation is 7) and since we have delocalization in coordinate
6 as well,
we end up with the semi-localized  M5-M5' brane system.

\begin{figure}[htb]
\begin{center}
\epsfxsize=2.5in\leavevmode\epsfbox{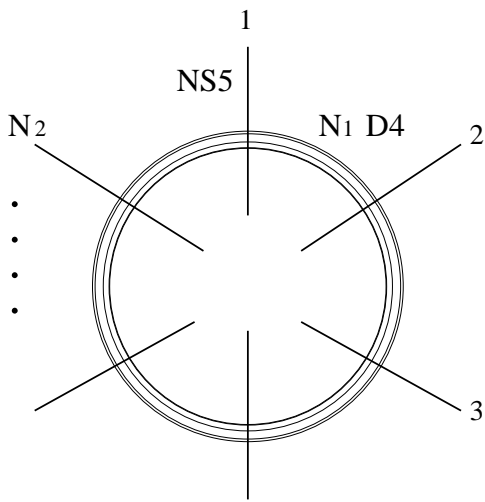}
\end{center}
\caption{ The Type IIA elliptic brane
system corresponding to the semi-localized M5-M5' brane configuration. It consists of
$N_2$ NS5-branes with worldvolume coordinates $(0,1,2,3,4,5)$ 
periodically arranged in the 6-direction and 
$N_1$ D4-branes with worldvolume coordinates $(0,1,2,3,6)$ 
stretched between them. 
}
\label{ell}
\end{figure}

The four dimensional theory at low energies on the
D4-branes worldvolume 
is an $SU(N_1)^{N_2}$ gauge theory with matter in the bi-fundamentals.
The metric (\ref{metric11}) can be viewed
as providing the 
eleven-dimensional supergravity description
of an M5 brane with worldvolume $R^4\times \Sigma$  \cite{witten}, where $\Sigma$  
is the 
Seiberg-Witten holomorphic curve (Riemann surface) associated with
the four dimensional SCFT at the origin of
the moduli space of vacua \cite{FS}.
In this brane set-up the R-symmetry group $SU(2)_R \times U(1)_R$ is realized as 
the rotation group $SU(2)_{8910} \times U(1)_{45}$. The dimensionless gauge coupling
of each $SU(N_1)$ part of the gauge group is $g_{YM}^2 \sim g_s N_2$.

The ten dimensional metric describing the elliptic Type IIA brane configuration
is  
\beq
ds^2_{10} = H^{-1/2}_4 dx_{||}^2+
H^{1/2}_4 d\vec{x}^2 
+H^{-1/2}_4H_{NS}dy^2+H^{1/2}_4H_{NS}d\vec{z}^2 \ ,
\label{dpns}
\eeq
where 
\begin{equation}
H_4=1+{{4\pi g_sl_s^4 N_1N_2}\over{[|\vec{x}|^2+
2 l_s N_2|\vec{z}|]^{2}}},\ \ \ 
H_{NS}={{l_sN_2}\over {2|\vec{z}|}} \ .
\label{dpnsharm}
\end{equation}
Its near-horizon limit is
\be
ds^2_{10}= l_s^2 \left (\frac{U^2}{R^2}dx^2_{||}+
R^2\frac{dU^2}{U^2}+ d\cM_5^2 \right ) \ ,
\ee
where
\beq 
d\cM_5^2 = 
R^2(d\alpha^2+\cos^2\alpha d\theta^2
+{\sin^2\alpha\over 4}d\Omega_2^2\bigg)
+{N_2^2\over {R^2 \sin^2\alpha}}dy^2 \ ,
\label{metric10}
\eeq
and $R^2=(4\pi g_sN_1N_2)^{1/2}$.
The curvature of the metric is 
${\cal R}\sim  \frac{1}{l_s^2R^2}$ and it has a dilaton 
$e^{\Phi}={{g_sN_2}\over {R}}\sin\alpha^{-1}$.
This supergravity solution can be trusted when both the curvature and the 
dilaton are small. Away from $\alpha=0$ we need large $N_2$ and 
$N_2 \ll N_1^{1/3}$. 
When the latter condition is not satisfied the dilaton is large and  we should consider
the eleven-dimensional description.

When both sets of M5 branes are fully localized the supergravity background
is not known.
The matrix description of intersecting M5 branes \cite{KOY} describes this system.
A discussion on  this case can be found in \cite{yoz}.

\section{BPS String Solitons}

The centrally extended $\cN=2$ supersymmetry algebra in four dimensions
takes the form \cite{FP}
\beqar
\{Q_{\alpha}^A, \bar{Q}_{\dot{\alpha}B} \} &=& \sigma_{\alpha\dot{\alpha}}^{\mu}P_{\mu}\delta_B^A
+ \sigma_{\alpha\dot{\alpha}}^{\mu}Z_{\mu B}^A,~~~A,B=1,2
\ , \nonumber\\
\{Q_{\alpha}^{A}, Q_{\beta}^B \} &=& \varepsilon_{\alpha\beta}Z^{[AB]} +
 \sigma_{\alpha\beta}^{\mu\nu}Z_{\mu\nu}^{(AB)} \ ,
\label{susy}
\eeqar
with $Z_{\mu A}^A = 0$.
The $SU(2)_R$ part of the R-symmetry group  acts on the indices $A,B$.
In addition to the particle charge $Z^{[AB]}$, there are in (\ref{susy})
the string charges $Z_{\mu B}^A$ in the adjoint representation of $SU(2)_R$  
and the membrane charges $Z_{\mu\nu}^{(AB)}$ in the 2-fold symmetric representation
of  $SU(2)_R$.
Thus,  $\cN=2$ supersymmetric gauge theories in four dimensions
can have BPS strings and BPS domain walls in addition
to the well studied BPS particles.

The BPS particles and BPS strings are realized by stretching M2 brane between the
M5 and M5' branes.
When the membrane worldvolume coordinates are $(0,x,y)$ where $x$ is one of the $\vec{x}$
components and $y$ is one of the $\vec{y}$ components we get a BPS particle
of the four dimensional theory.
It is charged under $U(1)_R$ that acts on $\vec{x}$ and is a singlet
under $SU(2)_R$ that acts on $\vec{z}$. 
The BPS particles
exist on the Coulomb branch of the gauge theory on which $U(1)_R$ acts, as in figure 2, 
and their mass is given by the Seiberg-Witten solution.

When the membrane worldvolume coordinates are $(0,1,z)$ where $z$ is one of the $\vec{z}$
components  we get a BPS string of the four dimensional theory.
It is not charged under $U(1)_R$ and it transforms in the adjoint 
$SU(2)_R$ since $\vec{z}$ transforms in this representation. The BPS strings
exist on the Higgs branch of the gauge theory on which $SU(2)_R$ acts, as in figure 2.
The tension of the BPS string is given by $\frac{|\vec{z}|}{l_p^3}$ where $|\vec{z}|$
is the distance between the M5 and M5' branes and it is finite in the field theory
limit.
It is easy to see that we do not have in this set-up BPS domain walls since stretching a
membrane with worldvolume coordinates $(0,1,2)$ breaks the supersymmetry completely.
One expects BPS domain wall configurations when the moduli space
of vacua has disconnected components.

In the Type IIA picture the BPS string is constructed by stretching a D2 brane between
the D4 branes and NS5-brane.
The seperation between these two types of branes in the $\vec{z}$ direction
has two interpretations  
depending on whether the four dimensional gauge group has
a $U(1)$ part or not \cite{GK}. If is does then the separation is interpreted as 
the $SU(2)_R$ triplet FI parameters
$\vec{\zeta}$. The  BPS string tension is proportional to $|\vec{\zeta}|$.
If the gauge group does not have a $U(1)$ part the  
seperation between these two types of branes in the $\vec{z}$ direction is 
interpreted as giving a vev to an $SU(2)_R$ triplet component
of the meson $\tilde{Q}Q$ which transforms under $SU(2)_R$ as 
${\bf 2} \times {\bf 2} = {\bf 3}\oplus{\bf 1}$.
The BPS string tension is proportional to this vacuum expectation value.
At the origin of the moduli space, where we have SCFT, the string becomes tensionless.

\begin{figure}[htb]
\begin{center}
\epsfxsize=3in\leavevmode\epsfbox{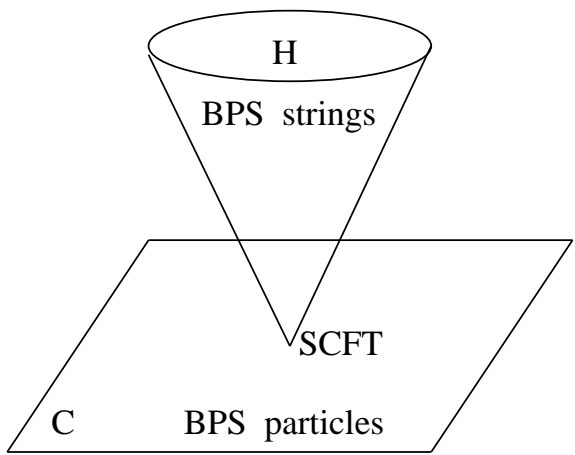}
\end{center}
\caption{There are BPS particles on the Coulomb branch and BPS
strings on the Higgs branch. The SCFT is at the intersection of these two branches. 
}
\label{ch}
\end{figure}

We can use the AdS/CFT correspondence in order to compute the self energy of 
a string and a potential between two such strings of opposite
orientation \cite{maldacena}. 
This is obtained, in the supergravity approximation, by a minimization
of the M2 brane action 
\be
S={1\over {(2\pi)^2 l^3_p}} \int d\tau d\omega d\sigma
\sqrt{det G_{\mu\nu}\partial_a x^{\mu}\partial_b X^{\nu}}
\ee 
in the background (\ref{metric11}), (\ref{M}).
Consider a static configuration $x^0=\tau,\,\, x^1=\omega$ and $x^i=x^i(\sigma)$.
Then, the energy per unit length of the string is given by
\be
E=\int{U\sin\alpha \over {4\pi^2 N_2}} \sqrt{dU^2 +U^2 d\alpha^2} \ .
\ee
Thus, the string self-energy 
\be 
E= {\cal E}(\alpha){U^2\over N_2} \sim {\cal E}(\alpha) \frac{N_1}{(\delta x_{||})^2} \ .
\label{se}
\ee
The function ${\cal E}$ parametrizes the dependence of the self-energy on the coordinate
$\alpha$. From the field theory point of view, this parametrizes 
a dependence on the moduli that parametrize the space of vacua. $\delta x_{||}$ is a cut-off.
In (\ref{se}) we used the
holographic relation between a distance  
$\delta x_{||}$ in field theory and the coordinate $U$ which in our case reads\footnote{
This correponds to the familiar relation $\delta x_{||} \sim \frac{(g_{YM}^2N_1)^{1/2}}{U}$
\cite{SW}.}
\be
\delta x_{||} \sim \frac{(N_2N_1)^{1/2}}{U}  \ .
\label{hol} 
\ee   

Similarly, the potential energy per units length
between two strings of opposite orientation seperated by distance $L$ is
\be
E \sim -{N_1\over L^2} \ .
\label{pot}
\ee
here $L$ is distance between two strings. 
The results (\ref{se}),(\ref{pot}) do not depend on $N_2$ which means that they
do not depend on the gauge coupling of the four dimensional 
field theory $g_{YM}^2 \sim N_2$. This is not unexpected.
The gauge coupling can be viewed as a vacuum expectation value  of a scalar
in the vector multiplet and does not appear in the hypermultiplet metric.
Since the BPS strings exist
on the Higgs branch it is natural that their potential and self-energy do not depend
on 
the gauge coupling. However, (\ref{se}) and (\ref{pot}) are only the large $N_1$ results
and will presumably have $1/N_1$ corrections.

As noted above, the theory on the intersection
of two sets of M5 branes containes BPS particles, which arise from
M2 branes stretched in the directions $(0,x,y)$.
Minimization of the M2 brane action in this case yields the potential between
these two such objects of opposite charge separated by distance $L$  
\be
V\sim {(N_1N_2)^{1/2}\over L} \ .
\ee
Again, this is expected since for the four dimensional field theory it reads as
$V\sim \frac{(g_{YM}^2N_1)^{1/2}}{L}$.

The 11-dimentional supergravity action goes like
\be
l_p^{-9}\int \sqrt{-g} {\cal R}\sim N_1^2 N_2
\ee
suggesting that the number of degrees of freedom goes like $N_1^2 N_2$. 
This is also deduced from the two-point function of the stress energy tensor.
This is, of course,  expected 
for $SU(N_1)^{N_2}$ gauge theory. 
It is curious to note that there is
similar growth ($N^3$) of the entropy for a system of $N$ {\it parallel}
M5 branes \cite{KT} \footnote{We thank A. Tseytlin for pointing this out.}.
The one loop
correction has the following form
$l_p^{-9}\int \sqrt{-g}l_p^6 {\cal R}^4\sim  N_2$
which is suppressed by $N_1^2$ compared to the tree level action
suggesting that the field theory has a $1/N_1$ expansion, which again is in agreement
with the field theory expectation.

The Type IIA background (\ref{metric10}) is T-dual to Type IIB on $AdS_5 \times S^5/Z_{N_2}$
\cite{FS}. The latter is the dual description
of the $Z_{N_2}$ orbifold of $\cN=4$ theory \cite{KS}. 
For instance the number of degrees of freedom can be understood as $c(\cN=4)/N_2 \sim N_1^2N_2$.
We can identify the spectrum of chiral primary operators with the supergravity
Kaluza-Klein excitations as analysed in \cite{OT,Gukov}.

\vskip 1.5cm

\section*{Acknowledgement}
We would like to thank J. Gauntlett, E. Rabinovici, S-J. Rey, A. Tseytlin, C. Vafa
and D. Youm for discussions.
M.A. is supported by the John Bell
scholarship from the world laboratory.

\newpage

\begingroup\raggedright\endgroup

\end{document}